%% file: main.tex
\documentclass[12pt,a4]{article}
\usepackage{epsfig}
 
\begin{document}
\bibliographystyle{alpha}
\pagenumbering{arabic}

\title{Structuring  Business  Metadata  in Data  Warehouse
Systems  for Effective Business Support}
\author{{\em N.L. Sarda}\\[1ex]
        Department of Computer Science and Engineering \\ Indian Institute of Technology Bombay \\ Mumbai, India \\[1ex]
        {\em nls@cse.iitb.ernet.in}}
\date{}

\maketitle
\thispagestyle{empty}

\subsection*{\centering Abstract}
\noindent
{ 
\small
 Large  organizations  today are being  served  by  different
types  of data processing and informations systems, ranging  from
the  operational (OLTP) systems, data warehouse systems, to  data
mining  and business intelligence applications.  It is  important
to  create an integrated repository of what these system  contain
and  do in order to use them collectively and  effectively.   The
repository contains metadata of source systems, data  warehouse,
and also the business metadata.  Decision support and  business  analysis
requires   extensive  and  in-depth  understanding  of   business
entities,  tasks,  rules  and the environment.   The  purpose  of
business metadata is to provide this understanding.
Realizing  the  importance  of  metadata,  many   standardization
efforts  has  been  initiated  to  define  metadata models.
In trying to define an integrated metadata and
information  systems for a banking application, we discover  some
important  limitations  or inadequacies  of the business  metadata
proposals.  They relate to providing
an  integrated  and  flexible  inter-operability  and  navigation
between  metadata  and  data, and to the  important  issue of
systematically  handling temporal characteristics and evolution  of
the metadata itself.  

In  this paper, we study the issue of  structuring  
business  metadata  so that it can  provide  a  context  for
business  management  and decision support when  integrated  with
data  warehousing.  We define temporal  object-oriented  business
metadata  model, and relate it both to the technical metadata 
and the data warehouse.  We also define  ways  of
accessing and navigating metadata in conjunction with data.
}\\\\
{\bf Keywords}
Data Warehouse, Metadata, Temporal database 
\newpage
\input{1paper.tex}

\input{2paper.tex}

\bibliography{main}
\end{document}

%% file: 1paper.tex
\section {Introduction}
     Large  organizations today are served by a large  number  of 
information  systems,  each  often built to  support  a  specific 
business  activity.   The operational systems, also  called  OLTP 
systems, are commonly built using commercial DBMS technology  for 
efficient    processing   of   business   transactions.     These 
distributed,    function-specific   information   systems    form 
information  `assets' of the organization.  The concept  of  data 
warehousing (DWH) has   rapidly  become  popular  to   leverage   the 
operational  data for strategic decision making across the  whole 
organization.   Data  warehousing is  a  technology  \cite{kimball, 
review}, that provides models, tools and techniques for  building 
integrated,  subject-wise data from the operational systems.   It 
provides  flexible  ways  to  access,  aggregate  and   visualize 
integrated   data  (possibly  from  across   different   business 
functions) based on selected parameters of interest, and navigate 
inside  the  data  `warehouse' so that  business  trends  can  be 
discovered and understood for better business decisions.

     In  the  context  of large number  of  information  systems, 
designed  and  maintained  by  different  organizational   units, 
operating  in heterogeneous computing environments, and  targeted 
towards  different business functions, it becomes essential  that 
ways  are found to correctly and consistently understand and 
co-relate  data from different sources in terms of formats,  meaning 
and  temporal  characteristics  of data.  For this,  we  need  to 
associate  comprehensive descriptions, called `metadata',  with 
the   data.   As  \cite{oim}  report  observes,   `Metadata,   or 
information  about data, has become the critical enabler for  the 
integrated management of the information assets of an enterprise'.

     \cite{oim}  defines  metadata  as  `descriptive  information 
about  the structure and meaning of data and of  the  applications 
and   processes   that  manipulate  data'.   The   metadata   has 
traditionally   been  classified  into  technical  and   business 
metadata.   The  technical  metadata 
specifies how exactly the data is structured and stored in  files 
or  databases.   This metadata allows applications and  tools  to 
access  and  manipulate the data.  The business  data,  expressed 
through   modeling  concepts  and  data  constraints,   help   in 
understanding  the  data and their usage.  In  order  to  achieve 
`strategic   advantages  of  on-line  access  to  all   knowledge 
maintained  in  the  distributed  computing  environment  of   an 
enterprise' \cite{oim}, a need for a centralized on-line metadata 
repository   has  been  recognized  by  researchers  (e.g.,   see 
\cite{dmdw99})  and standards organizations.  Such  a  repository 
needs   to  be  centrally  administered  for  effective  use   by 
the developers and end-users.

     While the technical metadata is relatively well-defined, the
business   metadata   has  posed  important   challenges 
due to the  diversity  of 
information that needs to be captured and modeled.  
(e.g., \cite{CWM}  does  not  cover business  metadata,  and  \cite{oim} 
standards  are still being proposed for addressing some  specific 
aspects  such as Business Engineering Model \cite{bem},  Business 
Rules  \cite{oim-brm}, and Knowledge Management model  \cite{oim-km}).

     The  Zachman Framework \cite{zach1}  established  different 
perspectives and dimensions for information systems within a 6 x  6 
table.   The  
perspectives cover directions and purpose (row 1), and the  nature 
of  business, including its structure, functions, organization,  etc. 
(row 2).  The dimensions include the `what',  `how', 
`where', `who', `when' and `why' aspects of a perspective. Thus, these 
two rows can be considered as defining  the business  metadata. The  row 
3 defines information system model using data models, process 
models  and  business  rules.   Considering  the  importance   of 
identifying and specifying the constraints of a business, the IBM 
user  Group,  GUIDE, published a standard for  defining  business 
rules  \cite{guide}.   The  report takes  an  information  system 
perspective,   where  business  rules  express   constraints   on 
creation,  updating and removal of persistent data.   The  report 
does not cover many other categories of issues (such as policies, 
guidelines,  processes, workflows, events, etc.).   The  business 
rules  conceptual model permits  us 
to  define  the business rules as  structural  assertions  (using 
`terms'  and  `facts'), action assertions (which  concern  system 
dynamics),  and  derivations.  The GUIDE model includes a
generic `policy' entity for relating rules to policies.  
     
     The  Business Engineering Model of OIM  \cite{bem}  proposes 
models for business goals (where goals are related to vision  and 
mission),   organizational  elements   (which   define 
resources and structure), business processes (which  describe  
business activities) and business  rules  (which  define 
business  constraints).   The  OIM  also  proposes  a   Knowledge 
Management  model  \cite{oim-km}  by which  an  organization  can 
catalogue and categorize information based on a suitable taxonomy 
of concepts. These various sub-models will permit users to interprete
data in the proper business context.
For  example,  an  information  system  may 
contain value for currency exchange rate.  This value needs to be 
related  to  the basic concepts of `currency',  `exchange  rate', 
time,  etc.  It also needs to be related to business goals  (such 
as  `ensure  stability  of exchange  rate'),  the  organizational 
elements  (such  as Exchange Control  Department) responsible  for 
(or, interested in) the goal, the related business processes,  and 
so on.


     As  we  move  from source application systems  to  the DWH 
applications, the granularity of metadata must also  be 
suitably  altered.  The operational systems process  transactions 
and perform activities at a detailed level, affecting  individual 
business entities.  The changes must be carried out as defined by 
business  rules.  
These rules specify detailed conditions and computations.  At the 
DWH application level, detailed business rules may not 
be  required.   It will be adequate to know  (in  a  higher-level 
articulation)  how  the  processing was done  and  what  are  the 
policies, so that they can be monitored and evaluated.

     The  business  metadata, like any other data,  changes  over 
time, since organizations change their missions, goals, policies, 
structure  and processes.  The business-level changes affect  the 
source  applications, which need to evolve to reflect changes  in 
data and processing.  The evolution of source applications, which 
feed  data  to the warehouse, must be captured in  the  warehouse 
metadata  so  that the subsequent analysis of  data  takes  these 
changes  into  account  in  evaluating  trends  and  taking   new 
decisions.   There  has been extensive  research  \cite{time}  in 
temporal databases, but none of the metadata standards explicitly 
define temporal properties for the metadata.

     While  the  motivation   for 
integration  is obvious, there have been a very few  papers  that 
suggest such integrated approaches.    
The  work by Stohr et al \cite{dmdw99} makes a  contribution 
in this direction.  They define a framework in which the business 
metadata  is  related  to  the technical  metadata  at  the  data 
warehouse  level.   Their  business model  is  somewhat  limited, 
as it is defined using a single "business concept" class.  The lowest 
level concepts may relate to either dimensions or measures at the 
warehouse  level.   They 
allows  DWH queries to be generated  while  navigating 
inside the business metadata.

     In defining our own framework for an integrated metadata and 
DWH environment, we make the following contributions:
\begin{itemize}
\item  We  structure  the  business metadata to  reflect  its  many 
     aspects, so that organizations can be guided to define  the 
     metadata  systematically and consistently.  We are guided in this both 
     by  the Zachman Framework and the Business Engineering submodel  of 
     OIM.
\item We  take into account the temporal aspects of metadata.  
     The  DWH applications invariably have Time as  an 
     important  dimension.  By providing `temporal  metadata', the 
     historical  data in DWH can be correctly  related 
     to other business changes.
\item We  provide an integrated and navigational access  to  data 
     from  metadata and vice-versa.  The navigation from data  to 
     metadata  is not defined in other approaches.  The navigation is 
     based on  the 
     associations between  various metadata and data entities, 
\end{itemize}
The   rest   of  the paper is organized   as   follows:   In 
Section  2,   we  give motivating examples  for  structuring  the 
metadata  and  for  metadata evolution (so  that  their  temporal 
characterization  can  be  done). We bring out different metadata
concepts, and also the requirements for metadata repository. In  
Section 3,  we  present   our 
integrated model for metadata and data systems. In Section 4, we give 
scenerios to illustrate exploration of data and metadata. Finally,
Section 5 contains conclusions.

\section{Characterizing Business Metadata}

     In   order   to support the various   organizational   roles 
such as operations, monitoring, control and decision-making,  the 
computer-based  data  processing  and  information  systems   are 
ubiquitous  today.  Invariably, an organization has  multiplicity 
of  such  systems,  each  with  a  specific  scope,  purpose  and 
functionality.  The data analysis  and 
decision support systems evaluate business performance against  a 
set  of goals, and facilitate new business decisions,  policies 
and goals. The DWH technology \cite{kimball} provides tools and 
techniques  for  efficiently organizing, browsing,  querying  and 
visualizing  large  amount of historical data  for  analysis  and 
decision-making.   The  data  in a DWH is  typically 
organized either as a `star schema' or a `multi-dimensional cube' 
consisting of {\it dimensions} and {\it facts or measures}.
The  measures may  represent   transactional 
data (aggregated, if necessary) or snapshot  data.
The data in operational systems go through 
the   `ETL  step',  where  the  relevant  data   are   extracted, 
transformed, co-related and finally loaded into the warehouse.

%
     In  today's   dynamic market conditions,  organizations  are 
placing  more  emphasis  on shorter decision  times  by  reducing 
organizational   hierarchies  and  empowering  their 
`knowledge-workers'. Consequently, a larger number of business users  need 
access   to  data  that  might  cut  across  the  boundaries   of 
organizational units and functions.  For them, the business  data 
needs  to  be  readily  accessible at  the  logical  level.   The 
technical   metadata,   primarily  meant  for   the   developers, 
administrators,  and the automated tools, is of little direct  use 
for business users.

     The   business   metadata, besides a   meaningful   
logical-level description of data, should also include the business 
context   of  data:  purpose,   relevance, 
potential  use, past usage, etc.  To some extent, such a  context 
can  be  provided  by  including in   business  metadata  the 
specifications  relating  to the first two rows  in  the  Zachman 
Framework. However, there  are  limited 
metadata  standards  efforts for these levels, and  possibly  not 
sufficient   experience   in  creating   comprehensive   metadata 
repositories at these levels.

     A  commendable    effort   in    building    a    high-level 
comprehensive business metadata has been undertaken at a  central 
bank in India.  The bank is involved in monitoring, planning  and 
regulating  of financial sector in the country.  It  proposes  to 
build  a  central DWH for  supporting  its  activities 
\cite{nag}.  The bank has a large number of departments, each with 
a specific  responsibility.   However,  the  basic  financial  data 
received from commercial banks and financial institutes is heavily 
cross-referenced.   Many financial parameters, such  as  exchange 
rate,  liquidity  ratio,  etc.,  have  ramifications  across 
functions, and are of interest to many departments.   The 
central  bank  also  has  to  prepare  extensive  statistics  for 
reviewing performance and ensuring  implementation of its  fiscal 
policies.  These statistics are used by the government for 
macro-economic planning,  and  by the bank  for  answering  important  
parliamentary questions.  The bank realizes the need not only for 
a  single  centralized  DWH, but also for  a  metadata 
repository to establish clear, consistent and complete meaning of 
data and other important business metadata \cite{nag}.

\subsection{Business Metadata  Categories}
     In  the  following, we discuss the nature of  metadata  with 
illustrating  examples  taken from the central Bank  case  study.  
However,  the categorization of the business metadata  is  fairly 
general,  and applies  to most businesses.  The categorization  can 
also be seen as an elaboration of the Zachman Framework.
\begin{itemize}
\item Functions: a broad classification of functions  or 
missions, such as Financial Supervision, Credit Management, 
Exchange Control, etc.
\item Organization Elements : define organizational  structure 
in  terms  of divisions, departments,  etc.   These  may 
include  Department  of Banking Supervision, Rural  Planning  and 
Credit   Department.
\item   Goal : a goal statement identifies purpose  for  an 
organizational element. These are often qualitative in nature. 
The monitoring and  control functions of the departments in the bank 
have explicit goals 
on  exchange rates, foreign exchange, interest rates, lending  to 
priority sectors, etc.
\item   Business  Entities  : an  organization  deals  with  many 
entities,  both  internal and external.  A description  of  these 
should  be  included in the metadata along  with  their  relevant 
attributes.  The examples could include different types of banks, 
non-banking   financial   institutes, and  business   sectors.
Entities may be  inter-related, 
and also associated with goal and other metadata categories.
\item Processes : these describe  at 
some  suitable level of details how the business  activities  are 
performed.   A process may be decomposed into sub-processes,  and 
be  associated  with other metadata such  as  functions, 
goal, business entity, etc.
\item   External Events : these are important happenings  (beyond 
the  control of the organization) which have impact  on 
some metadata, such as Policy and Functions.  They may be 
events  which  affect business entities  directly  or  indirectly 
(e.g., merger of two banks).
\item  Measures : these are quantitative parameters  that 
measure the effects of business activities.  This metadata type
specifies nature of measurement, the goals, business  entities,
and other metadata to  which they are related.
There  are  many important  measures  in  the  banking 
sector, such as deposits, credits, assets, incomes, expenditures, 
call money rates, etc.
\item    Evaluation  :  The  evaluation  class   represents 
organization's  evaluation  of business  measures  (as  obtained, 
after   suitable  filtering  and  aggregations,  from  the   data 
warehouse)  against the business goals.  The evaluations  may  be 
recorded  periodically.  They  act  as  records  of 
performance  of  the business.  These could be  simple  (such  as 
`satisfactory foreign reserves') or detailed (such as, `credit to 
agriculture sector in northern states is not growing at  expected 
rate').  The evaluation metadata would be associated  with 
organizational elements, goals and measures.
\item Action  : The  Action  metadata  class  permits 
recording  in  the  repository  any  specific  business   actions 
(possibly  as a consequence of evaluations and external  events).  
The  actions  would  relate  to evaluations.
\item  Business Concept : The various metadata classes  described 
above  can  be  generalized  as  business  concepts,  which,   in 
addition,  can also be used for introducing business  terminology 
(and  other miscellaneous business concepts not captured  by  the 
above categories).
\end{itemize}
\subsection{Metadata Characteristics}
     In  this  section,  we  highlight  important   metadata 
characteristics for  an  integrated   metadata 
environment.
\begin{itemize}
\item  Changes to Metadata : The  business  metadata, 
covered  under different categories in the previous subsection,  may 
change  over  time.   A  business user must  be  aware  of  these 
changes,   as   the  changes  may  alter  the   consistency   and 
comparability  of  data  over time.  The meaning  of  a  business 
concept  may change, or there may be changes in policies,  goals, 
and processes.  While  analyzing 
data  in  a  warehouse,  the data must  be  made  consistent  and 
comparable  (by  suitable transformations), if possible,  and the 
user  must  be  proactively made aware  of  changes  in  business 
metadata.  \cite{nag} gives many examples of metadata changes for 
the  central  bank.  For instance, the definition of  term  
`non-performing asset (NPA)'  has changed recently in terms of  duration 
of  non-payment  of a loan that leads to its classification  as 
NPA.

     Thus, the metadata are  temporal  in 
nature.   Each metadata has a time interval of  validity.  
This  interval  is  specified as beginning from  a  certain  time 
instant  (called,  FROM),  and ending  at  another  time  instant 
(called,  TO).   The description is effective in  the  real-world 
during  this  interval.   As  an implication,  the  data  in  the 
warehouse  must  also  be temporally characterized  so  that  the 
metadata and data can be temporally co-related.

\item  Metadata Abstraction Levels : The amount of metadata  that 
can be defined for an organization could be extensive, even if we 
assume  some  domain  background among  the  users.   Often,  the 
metadata   are  not  readily  available  (very   few 
organizations even have detailed 'operations and procedures' manuals).  
Moreover,  in  many  cases,  the  business  users  do  not   need 
definitions   of   primitive  concepts,  nor   the   details   of 
computations   and   processing.  Summary 
specifications are often adequate.  The detailed specifications
can be found, if  required,  in  the 
source code of some application. The business  metadata can provide 
a `drill-down' into the  application code for details of processes.

\item   Integrated Evolution : As business  requirements  change, 
the  application systems need to evolve.  The changes may affect 
database design as well as the  processing  logic.
The requirements change is naturally 
a consequence of changes in business policies, goals,  processes, 
rules,  etc.,  and should also result in  evolution  of  business 
metadata.  Also, the metadata repository must contain all  versions 
of  metadata with appropriate temporal validities.   To  minimize 
disruptions  and inconsistencies, organizations  should  plan 
integrated evolution of its metadata repositories and application 
systems.   An  evolution cycle may start with  business  metadata 
changes,  application  modifications,  and  end  with   capturing 
updates  to technical metadata.  When these steps  are  complete, 
the next ETL step will load data in DWH as per the  new 
business and technical metadata.  
\cite{ams}   presents  a  framework  for  application   evolution 
management.  It defines basic application components and  
their   temporal  characteristics. 

\item  Navigation Across Metadata and Data : An integrated system 
should  support a flexible access to both the metadata and  data.  
\cite{dmdw99}  propose such an environment where the  system  can 
generate  data  warehouse  queries while  navigating  within  the 
business   metadata.   We  propose  further  extension  to   this 
flexibility  :  it  should be possible to even go  from  data  to 
the temporally  consistent  metadata, see  metadata  changes,  record 
evaluations of business activities based on analysis of the data, 
and also record proposed business actions.
\end{itemize}
     Figure  \ref{fig1}  depicts an  integrated  environment  for 
metadata  management  and usage.  It shows linkages  between  the 
business and technical metadata, and the metadata evolution 
life-cycle.

%% file: 2paper.tex
\begin{figure}
\begin{center}
\mbox{\epsfig{file = 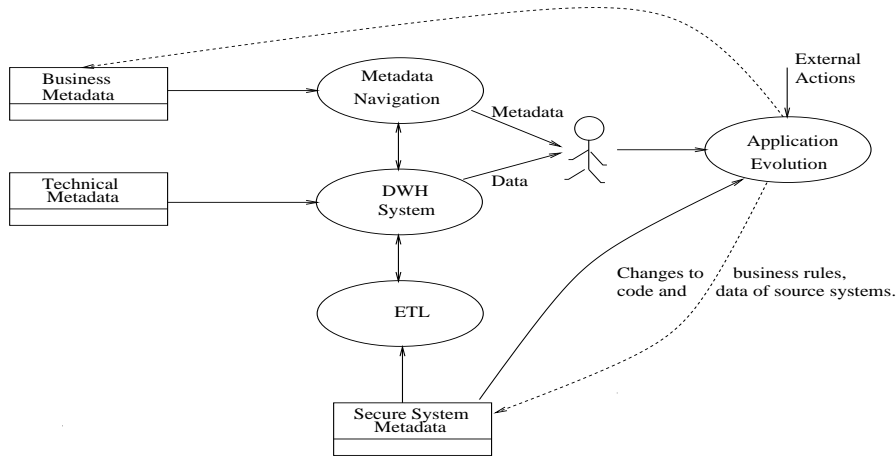, height = 6 cm, width = 12 cm}}
\caption{Integrated view of metadata management and usage}
\label{fig1}
\end{center}
\end{figure}
\section{Integrated Metadata Data Model}
\subsection{The UML Representation}
     In  the previous section, we presented different  categories 
of    business   metadata   and   also   highlighted    desirable 
characteristics  to  be supported by a metadata  repository.   In 
this section, we formally introduce the data model for metadata, 
using  {\it Unified Modeling Language}(UML)\cite{uml} as  the 
representation  language.  We use UML class diagram to  represent 
the business metadata classes and their inter-relationships.   In 
UML notation, rectangles represent classes, with class  names 
in the upper part and the main class attributes in the lower part 
(for clarity, we do not indicate class methods in the  diagrams).  
Inheritance relationship is shown using an arc with arrow-head 
pointing  to  the  super-class.    Associations   are 
represented by labeled arcs with an 'm' at the ends to 
represent  {\it many} type of associations.

\begin{figure}
\begin{center}
\mbox{\epsfig{file = 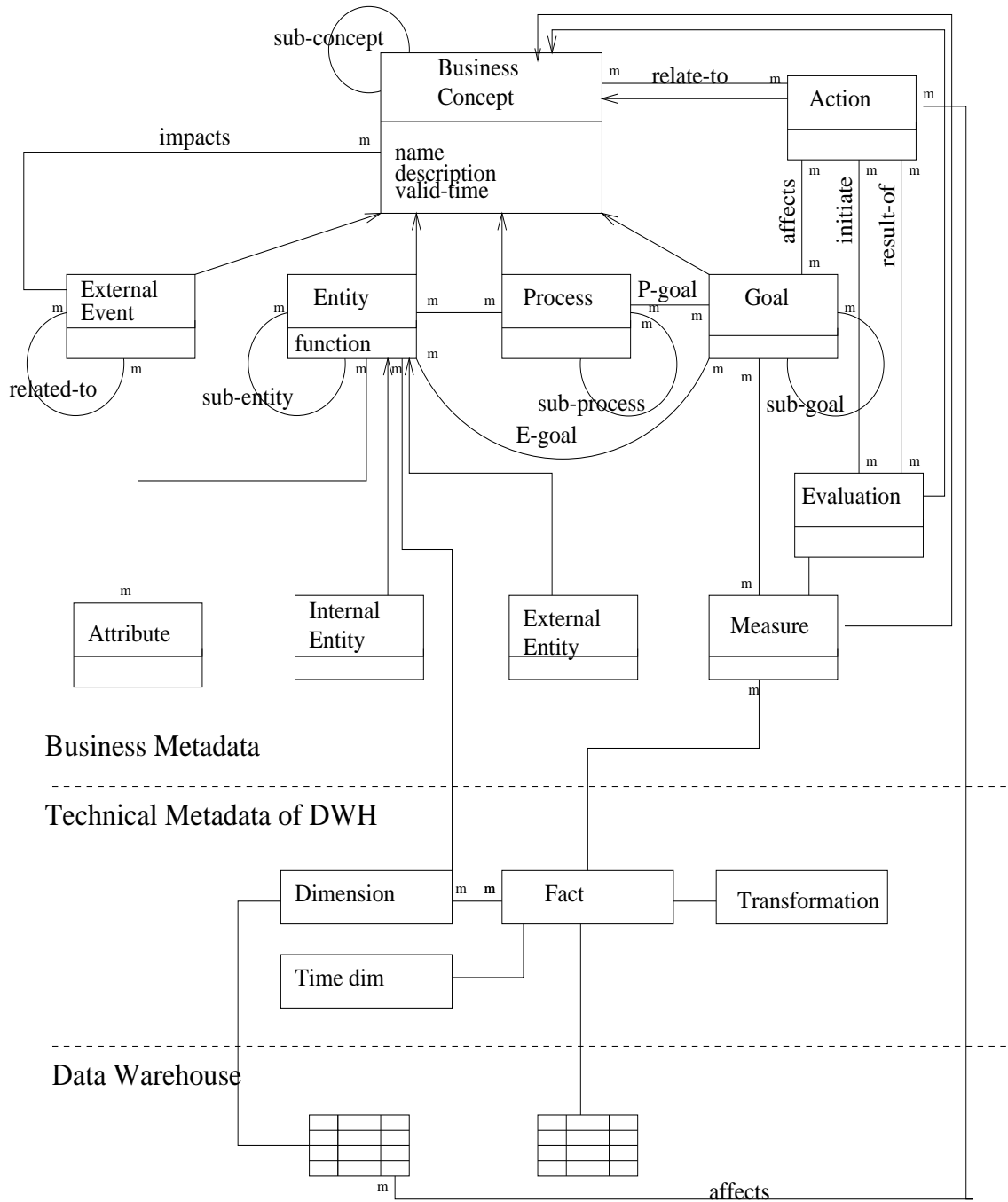, height = 18 cm, width = 15 cm}}
\caption{Business metadata model and integration with technical 
  metadata and DWH data}
\label{fig2}
\end{center}
\end{figure}

     The  overall metadata model is shown in  Figure  \ref{fig2}.  
The  figure, in fact, shows integration of the business  metadata 
with DWH technical metadata as well as warehouse data.  
The  associations across the boundaries of the three segments  in 
the  integrated  model represent the navigational  paths  between 
metadata  and  data  (in either direction).   In  the  technical 
metadata segment, we only show the primary metadata classes  used 
in structuring a warehouse with the purpose to establish important
relationships with business metadat.  It is not our intention to propose a 
new  model for technical metadata.  This  metadata  specification 
can use any of the proposed standards, such as OIM \cite{oim}  or 
CWM  \cite{CWM}.   The  OIM model, for  example,  uses  technical 
metadata classes such as Dimension, DimHierarchy, Cube,  Measure, 
etc.   OIM also allows specifications for transformations,  which 
can  describe how data from production databases  is  transformed 
before  loading data into the DWH.  
\\
\subsection{Metadata Classes and Associations}

     The  business metadata includes a generalized class  called 
Business  Concept.  Most other specific metadata classes  inherit 
from  this  class.   Its main attributes  are  concept  name  and 
description.  It also provides temporal validity attributes (FROM 
and  TO) to its subclasses and instances, so that  each  metadata 
definition  has  a  specific period of  validity.   The  temporal 
attributes  specify valid (i.e., real-world) time.  Thus, by  the 
accepted   definition   of   temporal   database   classification 
\cite{time}, the metadata repository is a `historical' database.  
This  class  may  have  instances  that  define  basic   business 
glossary,  and  that  metadata  which  do  not  fall  within  the 
classifications  given  by its sub-classes.  We  next  give  brief 
description   of   these  sub-classes  (which   were   introduced 
conceptually in the previous section):
\begin{itemize}
\item  External Event : It specifies some  real-world  happenings 
that  have  impact on the business in some  important  way.   Its 
instances  may represent both the descriptions of such events  as 
well   as  their  occurrences  (as  data  within   the   metadata 
repository; this data may not be captured in the DWH).  
This  class  has associations which indicate how the  events  are 
inter-related,  and  what  other business  aspects  these  events 
impact.  The valid-time primarily captures the definition time or 
occurrence time of these events.
\item  Entity : This class specifies various business  entities, 
which  may be either Internal or External.  The  Internal  entity 
occurrences   can   also   specify   business's    organizational 
elements.   Additional information about business entities  can 
be   captured   by   their  associations   with   the   Attribute 
specifications.  In a more general case, it may be appropriate to 
define association directly between BusinessConcept and Attribute 
classes  so that additional attributes can be defined  for  other 
metadata  classes also (in our case study, this  flexibility  was 
not  indicated).  The SubEntity association allows us  to  define 
hierarchies  among  business  entities.  For  example,  for  Bank 
entity,  we may define hierarchies to represent  `Foreign  Bank', 
`Nationalized Bank', `Rural Bank', etc.  The metadata  repository 
will  allow us to specify different business processes and  goals 
for these classes of entities.
\item  Process : The business process class  describes  business 
processes   and  their  subprocess  relationships  that   capture 
business  activities  and  tasks. 
\item  Goal  : The business goal class describes  objectives  and 
goals.   A  goal may have sub-goals.  Goals are also  related  to 
business  processes and entities through the P-goal  and 
E-goal associations.
\item  Measure : The measure class is used to  describe  business 
measures   that  are  used  to  evaluate   business   performance 
quantitatively.  A goal is associated with one or more  measures, 
and  a measure may relate to many goals.  A measure instance  may 
describe  the  computations  and  aggregations  required  to   be 
performed  on  data contained in a DWH to  understand  how 
achievement  of a business goal can be evaluated.   Although  not 
shown  in  the figure, the measure metadata class may  relate  to 
query or report entities in technical metadata.
\item  Evaluation  :  This  metadata  class  specifies  type   of 
evaluations  applicable  to  different Goal  metadata,  and  also 
evaluations  of measures at various times.  Its instances can  be 
associated  with Goal and Measure instances.  An  evaluation  may 
also  affect business entities and processes.  In the  model,  we 
capture   these  relationships  using  an   association   between 
Evaluation and the BusinessConcept classes.
\item  Action  :  This  class  represents  meaningful   business 
actions, and any specific actions undertaken as a consequence  of 
particular  evaluation.   We  define  separate  associations  to 
capture  evaluations  which  result  in  specific  actions,   and 
evaluations    which   are   consequences   of   some    actions.  
Relationships  of Action with other metadata are  captured 
by its association with the BusinessConcept super-class.
\end{itemize}

     The  business  metadata repository explicitly  captures  the 
various  dimensions  of Zachman framework  \cite{zach1},  and 
even  goes  beyond  the  framework  by  capturing  the   temporal 
characteristics of the metadata, and also providing for recording 
of  events,  evaluations  and  actions,  which  usually  have  no 
explicit  representation  in business information  systems.   The 
storage  of these data provides history of business  achievements 
and  decisions.   We may contrast this  detailed  structuring  of 
metadata  with  the  integraged model of  \cite{dmdw99}  where  a 
single Business Concept class is defined to capture business metadata.

\subsection{Linkages with Other Segments}

     The   business  metadata  segment  has  linkages  with   the 
technical  metadata and DWH segments of an  integrated 
metadata  and  data  environment.   The  important  linkages,  as 
envisaged in our model, are shown in Figure \ref{fig2}.

     The  business entities  in  the  business 
metadata  link  with  the  specification  of  dimensions  in  the 
technical  metadata  of  the DWH.   For  instance,  an 
(external)  entity Bank (with a suitable set of  attributes)  can 
map  onto a dimension called Bank.  The same dimension  may  also 
cover BankType entity to represent classifications of banks  into 
Nationalized,  Foreign,  Rural,  etc.   Similarly,  the  Business 
Concept  `Liability'  may  define  another  dimension.   We  have 
captured  this linkage with the Dimension metadata class  through 
the  association  with  BusinessConcept class.   We  do  not 
define  any explicit linkage to the Time Dimension.  There is  an 
implicit  linkage since the business metadata is  temporal.   The 
Time  Dimension represents the valid time line and  the  calendar 
defined in the temporal data model.

     Some  business metadata may actually represent the  data  in 
the  DWH itself.  For example, the  BankType  business 
entity  may have sub-entities NationalizedBank,  RuralBank,  etc.  
These  sub-entities  are also defined in  the  business  metadata 
repository.  However, these occur as rows in the dimension  table 
Bank  in  the DWH.  Another example:  the  `Liability' 
business concept may have sub-concepts like Deposits, Barrowings, 
etc.   These sub-concepts again appears as rows in the  Liability 
dimension  in the DWH.  To capture these linkages,  we 
define  an  association  from  BusinessConcept  to  rows  in  the 
dimension tables in the DWH.

     The  next linkage is between Measure  business  metadata 
class  and the Fact technical metadata class.   The  quantitative 
information specified by a Measure is available in the associated 
Fact table (for aggregation, filtering, etc.).

     Finally,  a  link is defined between  Action  metadata  and 
specific entities which are target of the action.  The  entities, 
in  general,  are  rows  in  a  dimensional  table  of  the 
DWH.   For  example,  an action might  indicate  that  some 
specific banks must reduce their assets in equities.

     As discussed in the next section, the metadata and  linkages 
across  the  three  segments  allow  effective  navigations   for 
understanding   of  business  situations  for   evaluations   and 
decision-making.

\subsection{Methods Supporting Navigations}

     The  various  metadata  classes have methods  for  not  only 
accessing  and  updating data in their instances,  but  also  for 
accessing   other  related  metadata  objects.    To   facilitate 
navigation,   we   have  defined  methods   to   follow   various 
associations  a class has.  We will illustrate  the  navigational 
methods for a few metadata classes.

     Consider  the  InternalEntity class.  It has  the  following 
methods:\\\\
{\it getSubEntity()}: to locate sub-entities of the entity\\
{\it getProcesses()}: to obtain process(es) in which the entity 
       is involved\\
{\it getGoals()}: to obtain goals related to the entity\\
{\it getAffectingEvents()}: to obtain external events that 
       affect the entity\\
{\it getActionsTaken()}: to obtain actions targeted towards the
       entity\\
{\it getDimension()}: to obtain entity instances (if applicable)
       stored in the rows of the dimension table.  This method
       navigates from business metadata to DWH.\\\\
For the Measure class, the following methods support navigation:\\\\
 {\it getGoals()} : to obtain goals which use the given measure\\
 {\it getEvaluation()} : to obtain evaluations based on the measure\\
 {\it getFacts()}  : to obtain fact descriptions  on  which  the 
       measure is based.\\\\

     Similar  kind of methods are defined for all the classes  in 
the  integrated  repository.   These methods  permit  going  from 
technical metadata to the business metadata and from data in  the 
DWH to the business metadata.  Temporal selection is  possible  with 
these methods by specifying time interval of interest.  Also, the 
method  {\it getHistory()} is provided for all temporal classes.   This 
method   can   be  used  to  obtain  all  earlier   states   (ie, 
descriptions)  of the metadata component.  The history, given  by 
the  past  states,  indicates evolution of  the  metadata.   In  a 
graphical  user interface, the methods are available as a  `drop-
down' menu by a mouse-click on a suitable class or instance.

\section{Exploring Metadata and Data}
     In  Section 2, we had highlighted importance  of  integrated 
environment for flexible access to metadata and data, as well  as 
access  to  the  business metadata  evolution  so  that  business 
activities and their performance can be understood in the context 
of  changing  business  structures,  processes  and  goals.   The 
integrated  model proposed in the previous section supports  both 
the requirements.  In this section, we briefly illustrate how the 
requirements are satisfied by the model.

\subsection{Scenerio 1 : Metadata to data to metadata}
     This  scenerio illustrates navigation within metadata,  then 
moving to the data in the DWH (transparently supported  via 
technical metadata by the warehouse tools) and returning back  to 
the  metadata.   By default, we access current instances  of  the 
metadata classes.  The steps in the scenerio could be as follows:
\begin{enumerate}
\item  Locate a department, an organizational element represented 
by  an  internal  entity (e.g.,  the  department  named  `Banking 
Supervision  Department').  We could go the subdivisions  of  the 
department, if required.
\item   Get  goals this entity is responsible  for  (through  its 
association with the Goal class).  There could be multiple  goals 
such as fraud detection, financial supervision, etc. 
\item   Given  a  goal,  locate how the  goals  are  measured  by 
following the Goal to Measure association.  There may be multiple 
measures  relevant  to a goal (alternatively, the goal  may  have 
subgoals).
\item   From a selected Measure, go the warehouse (cube  or  star 
schema)  that  records  values.   One  may  now  perform  various 
DWH operations like filtering, aggregations,  comparisons, 
etc.,  and  draw  a conclusion about how well the  goal  has  been 
achieved.  The conclusion can be stored as Evaluation instance in 
the metadata repository.
\item  From the measures visited earlier, move to other  goals 
that the measure is relevant for.
\end{enumerate}

\subsection {Scenerio 2: Metadata evolution}

     Since  the metadata as well as DWH data have  temporal 
validities,  the  integrated model permits us to  access  changes 
that  have  taken place in the business environment.  It  may  be 
noted  that  most  facts  recorded in  the  DWH have 
temporal  validities given by the corresponding  time  dimension.  
For  example,  the `Income Interest' is a  measure  recorded  for 
every  quarter for every bank.  We can observe how  this  measure 
has changed over time for a bank.  The metadata evolution, on the 
other hand, would show if the definition of the concept `Interest 
Income'  itself  has changed over time.  Let us  illustrate  this 
scenerio with a few steps:
\begin{enumerate}
\item  Access and examine the data in a DWH cube.
\item  Access the Measure metadata associated with the
cube.
\item  Access  the  goals associated with any  of  the  measures 
encountered  above.   Select one of the goals  (e.g.,  one  which 
prescribed  that  NPA (non-performing assets) be less  than  some 
percentage of gross assets).
\item   Display  NPA of Bank entities over time  (possibly  in  a 
convenient spread-sheet format).
\item  Choose an NPA value, and access its metadata description.
\item   For  the  NPA business concept, get  its  history  (i.e., 
evolution) over the last 2 years.  Note any major changes in  NPA 
definition  (which may explain the trend of NPA values seen  in  the 
earlier step).
\item   For  a selected entity, say XYZ Bank,  get  its  history.  
This may show changes in the bank's attribute values.
\item  For the selected bank, get external events that might have 
affected  it in the last 6 months.  This may, for  example,  show 
that it acquired another bank.
\end{enumerate}
     
     This  scenerio illustrates that the model allows  to  access 
changes in business metadata both at concepts level and at entity 
levels.   One may explore not only the DWH data,  but  also 
business environment changes.

\section{Conclusions}

     Large  organizations today need flexible access to various  kind 
of  information that is present in its operational systems.   The 
data  warehousing technology facilitates creation  of  integrated 
and  subject-wise  history data, and provides  flexible  ways  to 
access, aggregate and visualize the information in the DWH.  
In  order  for  users to know the meaning and  structure  of  the 
available  data,  we need to build `metadata  repositories'  that 
contain comprehensive descriptions of the data.  The metadata has 
been broadly classified into the business and technical metadata.  
It  is the business metadata which provides appropriate  business 
context  for  the  understanding and analysis  of  data  and  for 
decision-support.  Considering the importance of metadata,  major 
standardization  efforts, notably the OIM of MDC  \cite{oim}  and 
CWM of OMG \cite{CWM}, have been undertaken with primary focus on 
the  technical  metadata.   The  business  metadata  poses   many 
challenges  in terms of scope, abstractions and  structure.   The 
Zachman  framework \cite{zach1} has provided guidelines, and  the 
Business  Engineering Model proposals, covering  business  goals, 
organizational  elements, business processes, business rules  and 
knowledge management, of OIM emerge as a comprehensive effort  in 
this direction.

     In  an  attempt  to  build  an  integrated DWH 
environment  for  a  central bank, the  structuring  of  business 
metadata has merged as a major challenge \cite{nag}.  Unlike  the 
integrated model of \cite{dmdw99}, where all aspects of  business 
are  generalized  in  a single metadata  class,  called  Business 
Concept,  we considered it necessary to explicitly structure  the 
business  metadata  into many categories.  The  other  challenges 
included  changes to metadata itself over time, and  navigational 
access between metadata and data in either direction.

     The integrated metadata and DWH model proposed in 
this  paper  emerged as a solution for  the  bank's  requirements 
(although  it  is clear to us that a few  iterations,  after  the 
initial  implementation  and subsequent user feedback,  would  be 
necessary  to fine-tune it).  The model achieves  integration  of 
the three segments, business metadata, technical metadata and the 
DWH,  to allow the users  to  interpret,  understand, 
access and analyze the data in the DWH, taking into account 
the  temporal characteristics of both the metadata and data,  and 
then  evaluate  achievement  of business goals  and  plan  future 
business  actions.  These are the essential requirements for  the 
bank  for its regulatory and planning functions.  Like  the  most 
emerging  metadata  standards, we  use  object-oriented  modeling 
using  the UML representation.  We define temporal  validity  for 
all  classes,  and  navigational  methods  so  that  all  related 
metadata and data for any entity can be obtained.

     The  future work consists of fine-tuning the model based  on 
the  experience,  and  bring  it closer  to  the  OIM's  Business 
Engineering Model (BEM), but at the desired level of abstractions 
for  the application.  It may be noted that BEM in some  respects 
is too detailed (e.g., in the specification of business rules and 
processes),  and it does not support evolution and  changes.   We 
also plan to facilitate ad-hoc query support as in  \cite{dmdw99} 
from navigation and selections in the business metadata.